\documentclass[fleqn,usenatbib]{mnras}

\usepackage{newtxtext,newtxmath}

\usepackage[T1]{fontenc}

\DeclareRobustCommand{\VAN}[3]{#2}
\let\VANthebibliography\thebibliography
\def\thebibliography{\DeclareRobustCommand{\VAN}[3]{##3}\VANthebibliography}

\usepackage{graphicx}	
\usepackage{amsmath}	

\title{Search for pulsars in the summed power spectra at a frequency of 111 MHz}

\author[S. A. Tyul'bashev et al.]{
S.A. Tyul'bashev,$^{1}$\thanks{E-mail: serg@prao.ru (SAT)}
M.A. Kitaeva,$^{1}$
V.S. Tyul'bashev,$^{1}$
V.M. Malofeev,$^{1}$
G.E. Tyul'basheva$^{2}$
\\
$^{1}$ P.N. Lebedev Physical Institute of the Russian Academy of Sciences, Astro Space Center, Pushchino Radio Astronomy Observatory,\\
Radiotelescopnaya 1a, Moscow reg., Pushchino, 142290, Russia \\
$^{2}$ Institute of Mathematical Problems of Biology, brunch of Keldysh Institute of Applied Mathematics,\\  
Vitkevich 1, Moscow reg., Pushchino, 142290, Russia\\
}

\date{2019}

\pubyear{2019}

\begin{document}
\label{firstpage}
\pagerange{\pageref{firstpage}--\pageref{lastpage}}
\maketitle

\begin{abstract}
The search of pulsars in monitoring observations being carried out for 5 years using LPA LPI radio telescope was done in 96 spatial beams covering daily 17,000 square degrees. Five new pulsars were detected. Candidates into pulsars were selected in the summed power spectra. The use of a noise generator allowed to renormalize the data and to perform correct summing up of the power spectra for individual directions in the sky. Herewith, the sensitivity increased more than 20 times in comparison with individual sessions of observations. For pulsars with pulse durations greater than 100 ms at declinations $+30^o < \delta < +40^o$ it is equal to 1.2 mJy and 0.4 mJy in/out of the Galaxy plane.
\end{abstract}




\section{Introduction}

In the recent decade, there is a raising of activity in the search for pulsars. The reasons for this raising is due to the appearance of broadband recorders, the increasing of the speed of data processing, improvements in the algorithms for search of pulsars, followed by re-processing of archived records, the development of new search methods, the appear of new radio telescopes. Previously the search for pulsars was carried out repeatedly in the all celestial sphere. It makes sense to conduct new searches only on radio telescopes with high fluctuation sensitivity. That is why such programs are executed on the 300-meter Arecibo telescope (\citeauthor{Deneva2013}, \citeyear{Deneva2013}), 100-meter telescopes in Effelsberg and Green Bank (\citeauthor{Boyles2013} (\citeyear{Boyles2013}), \citeauthor{Barr2013} (\citeyear{Barr2013})), 64-meter telescopes in Parkes (\citeauthor{Keith2010}, \citeyear{Keith2010}), as well as on the aperture synthesis system LOFAR (\citeauthor{Coenen2014}, \citeyear{Coenen2014}), GMRT  (\citeauthor{Joshi2008} (\citeyear{Joshi2008}), \citeauthor{Bhattacharyya2018} (\citeyear{Bhattacharyya2018})). Pulsar search is also declared as one of the main objectives of the 500-meter telescope FAST (\citeauthor{Li2018}, \citeyear{Li2018}).

In 2013, at the Pushchino radio astronomy Observatory (PRAO) on Large Phased Array of the Lebedev Physical Institute (LPA LPI) after its complete upgrade, round-the-clock monitoring observations at 111 MHz were started in the framework of the Space Weather Project (\citeauthor{Shishov2016} (\citeyear{Shishov2016}). Despite the fact that the initial observations were carried out in only 6 frequency channals and with a sampling interval of 100 ms, the monitoring data can be also used to search for pulsars.

In the direct search of pulsars with the periods and the dispersion measure (DM) running over in the course of a 24-day monitoring at declinations $ +21^o < \delta < +42^o$, 7 new pulsars were found in individual records (\citeauthor{Tyulbashev2016}, \citeyear{Tyulbashev2016}). During the search for pulsars in the 4-year monitoring data at declinations $-9^o < \delta < +42^o$ using summed power spectra, 18 new pulsars were detected (\citeauthor{Tyulbashev2017}, \citeyear{Tyulbashev2017}).

In this paper, we present the results of our search for pulsars in 5-year interval of the monitoring data based on 6 frequency channel data.

\section{Observations and data processing}

Possibilities of upgraded LPA antenna, observation modes, digital recorders and data processing program are described in detail in the previous papers \citeauthor{Shishov2016}, (\citeyear{Shishov2016}), \citeauthor{Tyulbashev2016} (\citeyear{Tyulbashev2016}), \citeauthor{Tyulbashev2017} (\citeyear{Tyulbashev2017}). We’ll give here only a brief description.

The observations were done using meridian radio telescope LPA LPI. It is an antenna array made of wave dipoles with effective area of about 45,000 sq.m towards zenith. The zenith corresponds to the declination of $55^o$. The central frequency is 110.25~MHz and the operation frequency bandwidth is 2.5~MHz. The size of the space beam is approximately $0.5^o \times 1^o$, it allows to follow each source in the sky during 3-4 minutes per day. Currently, the monitoring part of the radio telescope in implemented in the 96 beams mode covering the sky in the declination range $-9^o <\delta < +42^o$. The area of about 17,000 sq. deg. can be seen in the sky daily. Continuous observations in the mode of 6 frequency channels (the bandwidth of single channal is 430 kHz) began in 2013. Since August 2014, observations have been conducted in parallel in the mode of 32 frequency channels at a sampling interval of 12.5 ms (the bandwidth of single channal is 78 kHz).

Monitoring observations are carried out by sessions during 1 hour. The beginning of each session is synchronized by the atomic frequency standard, and the time inside an hour block is measured by a quartz oscillator. The time error can reach 100 ms for an hour interval. As a result, the timing procedure for monitoring data has not yet been implemented.

The candidates for pulsars detected from 6-channel data are checked by 32-channel observations. After that, the pulsars are observed using a pulsar receiver with a high time and frequency resolution, which allows to reliably confirm the existence of the pulsar and to clarify its main characteristics (Malofeev et. al, in preparing).

At direct search of pulsars in separate sessions of observations, a normalization is not necessary, and it was not carried out (\citeauthor{Tyulbashev2016}, \citeyear{Tyulbashev2016}). When searching in the summed power spectra, self-calibration was done on the noise level in each frequency channel and for each day of observations (\citeauthor{Tyulbashev2017}, \citeyear{Tyulbashev2017}). The normalization procedure was carried out as follows: in each frequency channel on the time interval of 204.8 seconds (2048 points) the interference noise was removed, the baseline was subtracted and the noise level dispersion was calculated; the value of the signal amplitude at each point was divided by the dispersion, and as a result the final dispersion in the array became equal to one; the power spectra were calculated for each frequency channel independently, and then summed up; for a given direction in the sky, this procedure was carried out for each day of observation. It was expected that if on some days there was a difficult interference situation, or the sensitivity of the telescope fell for some other reason, the normalization procedure will suppress noise associated with interference, and when summing up power spectra for all days of observation, the overall deterioration in the summed power spectrum were negligible.

\begin{figure}
\begin{center}
	\includegraphics[width=0.9\columnwidth]{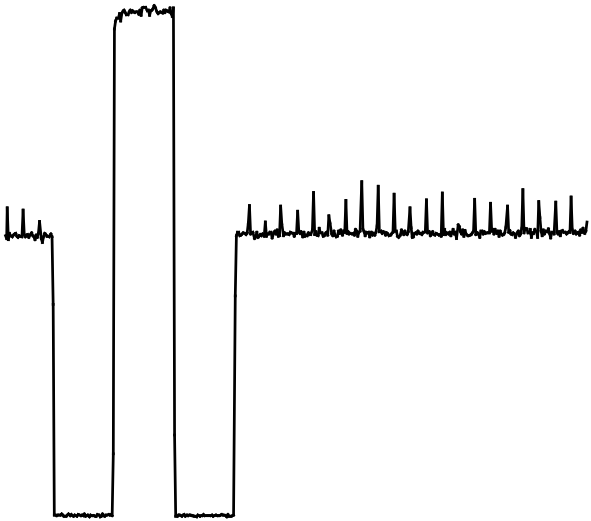}
    \caption{The figure shows the calibration signal and individual pulses of the known pulsar B1919+21 (PSR J1921+2153, P=1.337 s, DM=12.4 $pc/cm^3$), recorded in one of six frequency channels.}
    \label{fig:fig1}
\end{center}
\end{figure}

As shown by the analysis of the results obtained from early observations, even after the exclusion of obviously bad data, there are days of poor quality. Therefore, summing up all the remaining power spectra for the observed period, we invariably worsen the final signal-to-noise ratio (S/N) in the summed power spectrum compared to the spectrum, where data with poor quality would be excluded. In the present work, we have tested another data calibration system, using the noise generator signal. The calibration signal is recorded as "OFF-ON-OFF", where each part takes 5 seconds (see Fig.~\ref{fig:fig1}). The temperature of "ON" signal is 2400K, the temperature of "OFF" signal is about 300K (see more in \citeauthor{Shishov2016} (\citeyear{Shishov2016})).

During the day, the calibration signal is turned on 6 times, as a rule. The system amplification and therefore the dispersion of the digitized signal may change during the 4 hours between turning on of the calibration, for example, due to changes in ambient temperature. The system amplification may change as well from day to day. Therefore, for each direction in the sky, the height of the calibration signal before and after passing through the meridian of the pulsar candidate is estimated, and then using a linear approximation, the expected height of the calibration signal in the direction of the candidate is estimated in each frequency channel. All data were calibrated taking into account the amplitude of this signal equal to 2400K-300K=2100K.

\begin{figure}
\begin{center}
	\includegraphics[width=0.9\columnwidth]{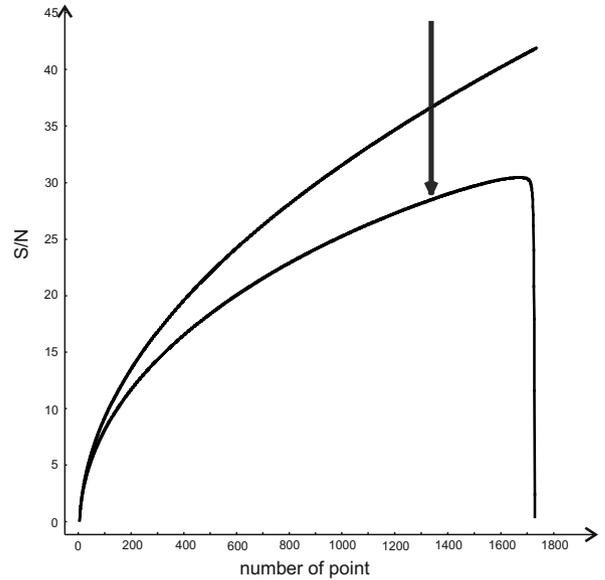}
    \caption{The marking on the vertical scale shows the S/N. The horizontal scale gives the number of observation days. The upper curve corresponds to the theoretical maximum possible signal-to-noise increase. The lower curve is the real behavior of the S/N in the summed power spectra. A sharp drop in the signal-to-noise ratio on the lower curve is associated with the days with the largest noise dispersion.}
    \label{fig:fig2}
\end{center}
\end{figure}

After the calibration of the data for each direction in the sky, it is possible to arrange the signal dispersions in order of their increase. The maximum dispersions will correspond to the low-quality observations and the minimum dispersions will correspond to the best-quality observations. The observation area is divided into approximately 40000 directions (pixels). For each pixel, after normalization using the calibration signal, the dispersion is calculated daily, and there are currently approximately $10^8$ noise level dispersion estimates.  Knowing noise dispersion for each day and the minimum dispersion over the observation interval, the real S/N can be estimated. While in the previous work (\citeauthor{Tyulbashev2017}, \citeyear{Tyulbashev2017}) it was assumed that the resulting increase in S/N after excluding the days with the low-quality of the observations is equal to the square root of the number of remaining days, now it is possible to test this assumption experimentally, taking into account the real estimates of noise dispersion. Fig.~\ref{fig:fig2} shows a typical picture of the expected and real increase in the S/N during observations for one of the directions in the sky.

As it can see from the lower curve, the S/N is first close to the theoretical curve and then farther and farther declines away from it. For five years of observations for different directions, from 30 to 60 days have been recorded with very low quality of the noise level. When ranking noise dispersion, these days fall into the end of the series and therefore the sum of all power spectra gives not an increase but a sharp drop in the S/N. There is a vertical line with an arrow on Fig.~\ref{fig:fig2}. It indicates the border of the number of days used for obtaining summed power spectrum in this pixel. This limitation was chosen for the reason that the S/N practically does not increase with further summation. In Fig.~\ref{fig:fig2} the summed power spectrum was calculated for 1300 days, and instead of the expected growth of S/N by 36 times the value of 28.7 times was obtained. The testing for different directions showed that the growth of S/N lies in the range from 20 to 30 times. Since the observations were carried out with the sampling interval of 100 ms, and the average half-width of the pulse of a pulsar is 20-30 ms, the final S/N will be additionally 1.5-2 times less.

\section{Results}

In previous work on the search for pulsars in the summed power spectra of LPA LPI radio telescope (\citeauthor{Tyulbashev2017}, \citeyear{Tyulbashev2017}), a number of criteria was used for additional confirmation of a new pulsar existence. These criteria were: a) the repeatability of the signal in sidereal time; b) the presence in the summed spectrum of at least two harmonics; c) the presence of a pronounced maximum on the dependence of the signal-to-noise ratio in the averaged profile of the dispersion measure; d) the existence of at least one record obtained in the 32- frequency channel mode and confirming the existence of a pulsar with an S/N observed in the averaged profile greater than 6; d) on a record with a double period, the averaged profiles should have approximately the same height.

For 21 pulsars out of 26 previously detected (\citeauthor{Tyulbashev2016} (\citeyear{Tyulbashev2016}), \citeauthor{Tyulbashev2017} (\citeyear{Tyulbashev2017}), \citeauthor{Tyulbashev2018} (\citeyear{Tyulbashev2018})), the regular observations have been made since July 2017 at the pulsar receiver with high frequency and time resolution ($4.88 kHz \times 460$ channels and time resolution 2.56 or 5.12 ms). Emission from 18 sources was confirmed, and observations for these sources are being made with the objective of clarify their coordinates and periods. For three, apparently weaker pulsars, the analysis is in process. The remaining 5 pulsars were preliminary identified as the weakest of the 18 pulsars previously found in the summed power spectra (\citeauthor{Tyulbashev2017}, \citeyear{Tyulbashev2017}). Currently, it is planned to conduct search observations for them with the pulsar receiver.

During the new processing of observations, the primary search for harmonics in the power spectrum was carried out using the BSA-Analytics software ({\it https://github.com/vtyulb/BSA-analytics}), but then all 40000 directions in the sky were checked visually. 87 candidates not analyzed in the early works were selected after screening out the known pulsars. Verification of the selected candidates showed that a significant part of them are known pulsars, observed in particular in the side lobes of the diagram of LPA LPI. There were 23 such pulsars. Eight candidates were periodic interferences of different nature. For the remaining 56 candidates, a search was conducted in the primary data, after which the 5 strongest sources with at least 2 harmonics in the summed power spectra (periods greater than 0.4s) were selected.

Verification of candidates was the same as in previous works. Of all the power spectra corresponding to the direction to the source, the spectra were selected, the harmonics of which were at the same frequencies as the harmonics in the summed power spectrum, and then the search was carried out with the variation of periods and dispersion measures. The date of observations, sidereal time of the middle of the checked interval and the expected period were set as input parameters. The program searches for a pulsar at an interval of $\pm 3$ minutes from a given time in increments of 20s to clarify the coordinates of the right ascension. Thus, the total length of the time interval at which the search is performed in the primary data is 6 minutes. At each step, the variation of periods was done within $\pm 10\%$ of the specified period. At each interval, the dispersion measures in the range $0-200 pc/cm^3$ are running over and the averaged profiles obtained on double periods with S/N greater than 5 are fixed. The processing results for each selected day were saved. It allows subsequently to load several processed days into the program and carry out both the summation of the averaged profiles and the summation of the S/N dependence on the dispersion measure being checked.

\begin{figure}
\begin{center}
	\includegraphics[width=0.9\columnwidth]{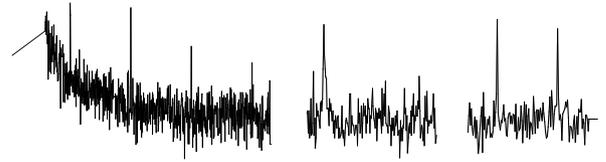}
    \caption{From left to right: summed power spectrum on which there is 4 harmonics, S/N dependence from DM (DM on the figure are from 0 to 200 $pc/cm^3$) and the averaged profile of J0305+1127 pulsar summed up with double period.}
    \label{fig:fig3_finish}
\end{center}
\end{figure}

Fig.~\ref{fig:fig3_finish} gives an example of power spectrum, S/N dependence from DM and assembled a double profile for the source J0305+1127. The summed power spectrum contains strong low-frequency noise, which is partially removed in the figure. The maximum of S/N dependence on DM falls by $26.5 pc/cm^3$.

For other pulsars found, the ratio of pulse duration to period is about the same, so all the obtained averaged profiles are very similar. We do not give them in this work, but posted them on the website ({\it https://bsa-analytics.prao.ru/en/pulsars/new/})together with the average profiles of previously discovered pulsars.

The results of the examination of 5 candidates, for which we obtained average profiles and estimated the dispersion measures on 32-channel data, are placed in the Table~\ref{tab:tab1}. The first column of the table shows the name of the pulsar in the J2000 agreement, the second and third columns give the coordinates of the pulsar right ascension and declination for the year 2000. The typical accuracy of the coordinates determination on the right ascension is $\pm 30s$, and on declination it is $\pm 15^\prime$. The fourth column gives the pulsar period determined with an accuracy of $\pm 0.0005s$, the fifth gives its dispersion measure, the sixth gives the visible half-width of the averaged profile. The half-width of the averaged profile can have large errors, since no account was taken of the possible broadening of the pulse beyond DM in the bandwidth of one frequency channel. The actual width of the averaged profile may be less than the estimate given.

\begin{table}
	\centering
	\caption{Characteristics of discovered pulsars.}
	\label{tab:example_table}
	\begin{tabular}{cccccc}
		\hline
		
$Pulsar name$ & $\alpha_{2000}$ & $\delta_{2000}$ & P & DM & $W_e $\\
		\hline
J0305+1127 & $03^h05^m50^s$ & $11^o27^\prime$ & 0.8636 & $26.5\pm 1.5$ & 16\\
J0350+2341 & 03 50 03 & 23 41 & 2.4212 & $61\pm 1.5$ & 21\\
J1740+2728 & 17 40 17 & 27 28 & 1.0582 & $35\pm 2$ & 21\\
J1958+2213 & 19 58 34 & 22 13 & 1.0502 & $85\pm 3$ & 21\\
J2210+2117 & 22 10 15 & 21 17 & 1.7769 & $45\pm 2$ & 25\\
		\hline
	\end{tabular}
	\label{tab:tab1}
\end{table}

\begin{figure}
\begin{center}
	\includegraphics[width=0.9\columnwidth]{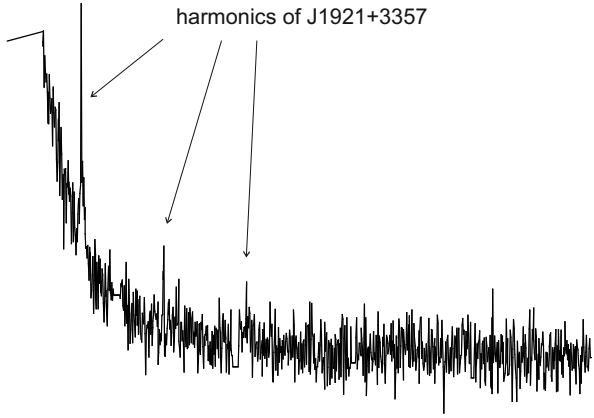}
    \caption{The summed power spectrum of J1921+3357. Harmonics multiples of one second and associated with internal interferences are removed from the spectrum.}
    \label{fig:fig4_finish}
\end{center}
\end{figure}

For 51 of the candidates it was not possible to find separate days on which the 32-channel data would allow to do a check up of candidates. An example of the strongest of these candidates is shown in Fig.~\ref{fig:fig4_finish}. For the source J1921+3357, three harmonics are visible, it is observed in two adjacent beams of the antenna. The period corresponding to the inverse frequency of the first harmonic is P=1.441s. The accuracy of the period determination is $\pm 0.005s$. The coordinate precision on the right ascension is $\pm 2^m$ and on declination it is $\pm 20^\prime$. Within the coordinates errors there is no identification with known ATNF pulsars ({\it http://www.atnf.csiro.au/people/pulsar/psrcat/}). There are 7 pulsars in the ATNF within the period measurement errors: J0754+3231 (P=1.442s; DM=39.99 $pc/cm^3$); J0934-5249 (P=1.445s; DM=100 $pc/cm^3$); J1703-4851 (P=1.396 s; DM=150 $pc/cm^3$); J1824-1350 (P=1.396 s; DM=551 $pc/cm^3$); J1910+1231 (P=1.441s; DM=258.6 $pc/cm^3$); J2010+3230 (P=1.442s; DM=371.8 $pc/cm^3$); J2317+2149 (P=1.444s; DM=20.87 $pc/cm^3$). The pulsars J0934-5249, J1703-4851, J1824-1350 can’t generate the side lobe due to the low declinations not reachable for observations by the LPA LPI. The pulsars J1910+1231 and J2010+3230 have high dispersion measure values and are not detected even in the main beams, and, moreover, will not display themselves in the side lobes. The pulsar J0754+3231 has a close declination and is visible in the summed power spectra, but its first harmonic is comparable in height to the first harmonic of our candidate, so the contribution to the far side lobe of the LPA LPI diagram by right ascension is unlikely. The harmonics of pulsar J2317+2149 have a very large S/N, and it is visible as well in the side lobes of the LPA, but we are not aware of the side lobes in the direction of the candidate for pulsars. At the same time, for J1921+3357 pulsar not a single individual power spectrum was found that could be used for obtaining estimates of the pulsar parameters. We have no doubt that this object is a pulsar, but the sensitivity of the LPA LPI is not sufficient for its verification. There is the similar situation with the other candidates for pulsars. In the paragraph "Discussion of results", we consider the possibility of confirming these candidates as pulsars.

In addition to the new pulsars, there are more than 100 known pulsars in the summed power spectra with periods $P>0.4s$ published in the ATNF and in our early works. The list of discovered known pulsars is constantly updated. These pulsars are not the subject of this work. Their averaged profiles and some additional information are posted on the website {\it https://bsa-analytics.prao.ru/en/pulsars/known/}.

\section{Discussion of result}

As noted in the previous paragraph, objects are observed in the summed power spectra that can’t be confirmed as new pulsars. Some of them may be noise or distant side lobes of known pulsars. However, among them must be also new pulsars. An example of such a pulsar is shown on fig.4 in the previous paragraph. Candidates for pulsars, as well as detected pulsars are observed in one or two adjacent beams, but the sensitivity of the LPA LPI radio telescope happens to be not sufficient to have a obvious maximum in one observation session of the S/N dependence on DM. In the previous work, extremely weak pulsar flux densities observed in the power spectra were already estimated. These estimates were 0.2 mJy outside the Galactic plane and 0.6 mJy in the Galactic plane at a frequency of 111 MHz (\citeauthor{Tyulbashev2017}, \citeyear{Tyulbashev2017}). Taking into account the calculated curves of the real growth of the S/N in the summed spectra (similar with fig.2), these estimates can be adjusted to 0.3 mJy outside the Galactic plane and to 0.9 mJy in the Galactic plane on the 4 years observation interval. Let us remind that in the work (\citeauthor{Tyulbashev2017}, \citeyear{Tyulbashev2017}), the estimation of sensitivity is given to the direction of the Zenith, that is, it is estimates of the maximum possible sensitivity. It was also said that the difference between the maximum and minimum sensitivity can reach one order of magnitude due to the particular features of the diffraction array. In our observations, the maximum declination was $42^o$, while the direction to the Zenith corresponds to the declination of $55^o$. Sources usually have coordinates that fall on the directions between the beams, which reduces the sensitivity of the search. It’s possible to make a rough estimation of the real sensitivity, taking into account the location of the beams on the declinations. It is approximately equal to 0.4 mJy and 1.2 mJy for declinations in the range of $+30^o < \delta < +40^o$ for directions out/in of the Galactic plane. For low declination in the range of $-9^o < \delta < +3^o$ in addition to the corrections for the cosine of the Zenith distance, the sensitivity falls due to the reduction of the effective bandwidth and is 1.2/3.6 mJy for directions out/in of the Galactic plane.

It is possible to make rough estimates of the limiting sensitivity of the upgraded LPA LPI antenna in a single session of observations of known pulsars, based on the measured pulsar flux density at 102 MHz (\citeauthor{Malofeev2000}, \citeyear{Malofeev2000}) and the observed signal-to-noise ratio in the averaged profile. These estimates allow us to hope for the detection of pulsars with a flux density of about 5 mJy in a single observation session. Formal estimates of the sensitivity based on the effective area and other known quantities give the limiting observations sensitivity in a single session of 4.4 mJy for sources outside the Galaxy plane in directions close to the Zenith (\citeauthor{Tyulbashev2016} (\citeyear{Tyulbashev2016}).

To confirm the new pulsars with flux density less than 5 mJy, extra efforts are needed. The direct way is observations with radio telescopes that have a higher sensitivity than LPA LPI. Presumably, such observations on the LOFAR core are most advantageous, since the smaller effective area of this antenna is compensated by wide bandwidts and the possibility of tracking. At the same time, the central frequency equal to 140 MHz, is close to our central observation frequency. Observations on the FAST radio telescope are even more advantageous because of its extremely high sensitivity achieved at small observation time intervals. A possible way out is also to make observations during long time intervals with our antenna. Since pulsars are variable objects, there is always the possibility that, with an average flux density of less than 5 mJy, they may have a higher flow density in single sessions.

Indirect way is the use of monitoring data from LPA LPI recorded in 32-channel mode. Processing of the observation data is carried out on home computers, and takes a long time. As mentioned in the paragraph "Observations and data processing", to reduce the data processing time, the power spectra were calculated for each pixel independently in each frequency channel for a given day, and then summed up. As the next step, the spectra were summed up for all observation days. In this operation information not only on the phase of the pulses, but also on DM was lost. However, information on the dispersion measure can be recovered, if for the original data for each dispersion measure to make preliminary shifts of the arrays in all frequency channels to compensate the expected dispersion measure, to sum up all the channels and then to calculate the power spectrum. In this case the harmonics will appear only in those spectra where the signal was summed up on the "correct" dispersion measure. If to collect these power spectra and sum them up for all the days, then the correct dispersion measures can be when the harmonics have a maximum height in the summed power spectrum. This method of extracting the dispersion measure for single records was considered, for example, in the Lorimer\&Kramer handbook \citeauthor{Lorimer2004} (\citeyear{Lorimer2004}. Evaluation of the dispersion measure different from zero, despite the absence of an averaged profile, will be a good confirmation of a pulsar existence. Currently, 32-channel data are being tested for known pulsars (\citeauthor{Malofeev2018}, \citeyear{Malofeev2018}). Processing of the monitoring observation data on the 4-year interval is planned.

It is interesting to note that in addition to the power spectra in which two or more harmonics are observed, there are also hundreds of single harmonics in the summed power spectra. Part of these harmonics is repeated in many beams and is associated, apparently, with industrial and internal interferences. However, a part of the harmonics is observed only in separate beams, and periods corresponding to these harmonics are not repeated.

Let us consider the possible nature of these harmonics. First, in the power spectra of pulsars, the first harmonic has, as a rule, a maximum height, and then the height of the harmonics decreases. Therefore, if the first observed harmonic has a small height, then the next harmonic may not be visible in the noise. Such pulsars are weak objects for observation by LPA LPI, and it is necessary to improve the data processing methods for such pulsars. Second, pure sinusoidal signals, that can be inherent to industrial interference, will make a single harmonic. The same single harmonic will be generated by such pulsars as alignment rotators, as is also pulsars, which due to the large dispersion measure have a signal broadening, and their pulses occupy most part of the period. The averaged profiles of these pulsars look unconvincing. For such pulsars, it is difficult to determine the S/N in the averaged profile, and there are large errors of determination of dispersion measures (see, for example, pulsar J1844+4117 (\citeauthor{Tyulbashev2017}, \citeyear{Tyulbashev2017}).

\begin{figure}
\begin{center}
	\includegraphics[width=0.9\columnwidth]{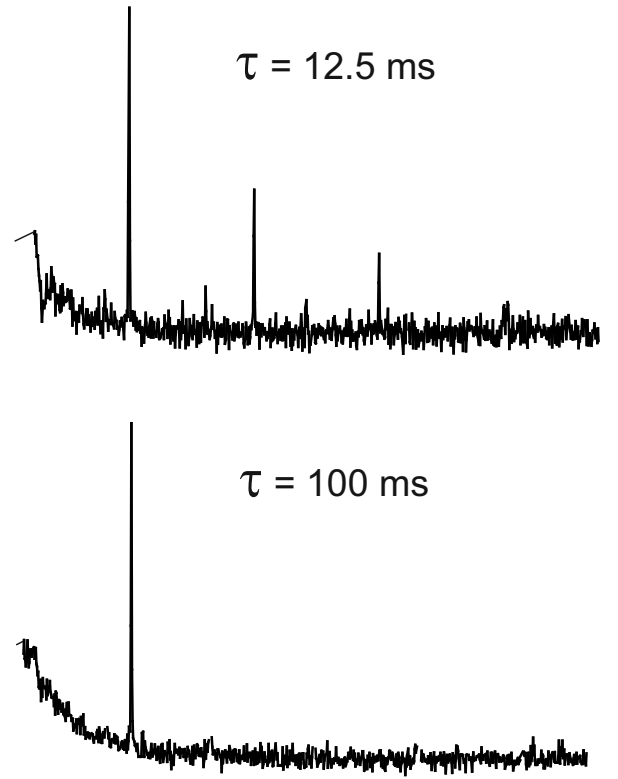}
    \caption{Power spectra of pulsar J1922+2110 from 32-frequency channel (top picture) and 6-frequency channel (bottom picture) data. At the beginning of the recordings, low-frequency noise is seen, which is partially removed in the pictures. The sampling times in the observation sessions is indicated at the power spectra.}
    \label{fig:fig5_finish}
\end{center}
\end{figure}

A partial solution to the problem of searching for pulsars in the spectra with one observed harmonic is the transition to 32-channel data. Fig.~\ref{fig:fig5_finish} shows two summed power spectra obtained from data recorded in the full 2.5 MHz band in the 6 and 32 frequency channel mode for the known pulsar J1922+2110. This pulsar has a dispersion measure DM=217 $pc/cm^3$ and the period P=1.0779s. Only the first harmonic is distinguished in the power spectrum from the 6-channel data. The 32-channel data shows 4 harmonics. On the power spectra shown in the Fig.~\ref{fig:fig5_finish}, the harmonics not related to the pulsar J1922+2110 are removed. The effect of the increase in the number of the observed harmonics probably reflects a narrowing of the observed pulses due to recording at a significantly higher sampling interval  of 12.5 ms instead of 100 ms.

This example shows that it is necessary to further develop the method of new weak pulsars detection and extraction of their parameters.

\begin{figure}
\begin{center}
	\includegraphics[width=0.9\columnwidth]{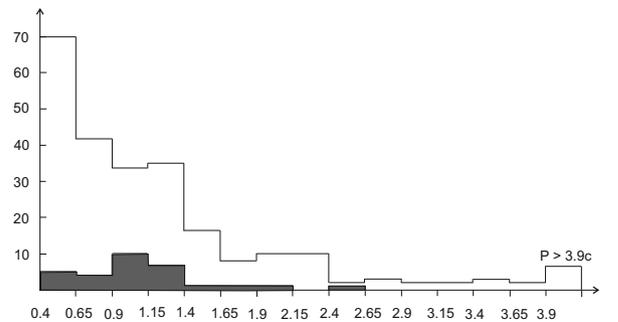}
    \caption{Histogram of ANTF pulsars periods distribution (not shaded part of the histogram). Within this histogram there is the histogram of periods distribution for pulsars detected at 111 MHz (shaded with black color)}
    \label{fig:fig6_finish}
\end{center}
\end{figure}

We have compared our pulsars with known pulsars from the ATNF catalog. To do this, pulsars that fall on declinations into the observation area ($-9^o< \delta < +42^o$), having dispersion measures $DM < 100 pc/cm^3$ and rotation periods P>0.4 s, were selected in the ATNF catalog. The limitations are related to the possibilities of our observation. Fig.~\ref{fig:fig6_finish} shows histograms of the distribution by periods for 243 ATNF pulsars and for 30 pulsars discovered by us (\citeauthor{Tyulbashev2016} (\citeyear{Tyulbashev2016}), \citeauthor{Tyulbashev2017} (\citeyear{Tyulbashev2017})), this paper).

The number of pulsars detected by us is small, but it is clear that the distributions are different. The striking feature is the deficit of pulsars with periods less than 0.9 s for the lower curve and the presence of the maximum in the range of 0.9-1.4 s. However, on the upper curve too, there is a small excess of pulsars in this range. It is possible that the pulsar deficit is associated with a lower search frequency (111 MHz), but most likely it reflects the low time resolution of our search (100~ms). This deficit is likely to decrease with the inclusion in the processing of data obtained from 32 frequency channels with more high time resolution (12.5~ms).

\section{Conclusion}

The search for pulsars in 6-frequency channel data in the 5 years interval has been made. Five new pulsars had been detected. Together with the results of the previous works  (\citeauthor{Tyulbashev2016} (\citeyear{Tyulbashev2016}), \citeauthor{Tyulbashev2017} (\citeyear{Tyulbashev2017}), \citeauthor{Tyulbashev2018} (\citeyear{Tyulbashev2018})), the total number of found pulsars increased up to 31. The search was also carried out for the 51 candidates for pulsars having in the summed power spectra not less than two harmonics. For those we failed to find any day on which it would be possible to specify their coordinates, periods, and dispersion measures. The flux density from these pulsars is expected to be less than 5 mJy at 111 MHz. It is demonstrated that when searching using the summed power spectra in the 5-year interval, the signal-to-noise ratio increases 10-30 times depending on the direction in the sky.

\bsp	
\label{lastpage}

\begin{thebibliography}{99}
\bibliographystyle{unsrt} 

\bibitem [Deneva(2013)]{Deneva2013} Deneva J.~S., Stovall K., McLaughlin M.~A., Bates S.~D., Freire P.~C.~C., Martinez J.~G., Jenet F., et al., 2013, ApJ, 775, 51. doi:10.1088/0004-637X/775/1/51

\bibitem [Boyles(2013)]{Boyles2013} Boyles J., Lynch R.~S., Ransom S.~M., Stairs I.~H., Lorimer D.~R., McLaughlin M.~A., Hessels J.~W.~T., et al., 2013, ApJ, 763, 80. doi:10.1088/0004-637X/763/2/80

\bibitem [Barr(2013)]{Barr2013} Barr E.~D., Champion D.~J., Kramer M., Eatough R.~P., Freire P.~C.~C., Karuppusamy R., Lee K.~J., et al., 2013, MNRAS, 435, 2234. doi:10.1093/mnras/stt1440

\bibitem [Keith(2010)]{Keith2010} Keith M.~J., Jameson A., van Straten W., Bailes M., Johnston S., Kramer M., Possenti A., et al., 2010, MNRAS, 409, 619. doi:10.1111/j.1365-2966.2010.17325.x

\bibitem [Coenen(2014)]{Coenen2014} Coenen T., van Leeuwen J., Hessels J.~W.~T., Stappers B.~W., Kondratiev V.~I., Alexov A., Breton R.~P., et al., 2014, A\&A, 570, A60. doi:10.1051/0004-6361/201424495

\bibitem [Joshi(2008)]{Joshi2008} Joshi B.~C., McLaughlin M.~A., Kramer M., Lyne A.~G., Lorimer D.~R., Ludovici D.~A., Davies M., et al., 2008, AIPC, 983, 616. doi:10.1063/1.2900311

\bibitem [Bhattacharyya(2018)]{Bhattacharyya2018} Bhattacharyya B., Lyne A.~G., Stappers B.~W., Weltevrede P., Keane E.~F., McLaughlin M.~A., Kramer M., et al., 2018, MNRAS, 477, 4090. doi:10.1093/mnras/sty923

\bibitem [Li(2018)]{Li2018} Li D., Wang P., Qian L., Krco M., Jiang P., Yue Y., Jin C., et al., 2018, IMMag, 19, 112. doi:10.1109/MMM.2018.2802178

\bibitem [Shishov(2016)]{Shishov2016} Shishov V.~I., Chashei I.~V., Oreshko V.~V., Logvinenko S.~V., Tyul'bashev S.~A., Subaev I.~A., Svidskii P.~M., et al., 2016, ARep, 60, 1067. doi:10.1134/S1063772916110068

\bibitem [Tyulbashev(2016)]{Tyulbashev2016} Tyul'bashev S.~A., Tyul'bashev V.~S., Oreshko V.~V., Logvinenko S.~V., 2016, ARep, 60, 220. doi:10.1134/S1063772916020128

\bibitem [Tyulbashev(2017)]{Tyulbashev2017} Tyul'bashev S.~A., Tyul'bashev V.~S., Kitaeva M.~A., Chernyshova A.~I., Malofeev V.~M., Chashei I.~V., Shishov V.~I., et al., 2017, ARep, 61, 848. doi:10.1134/S1063772917100109

\bibitem [Tyulbashev(2018)]{Tyulbashev2018} Tyul'bashev S.~A., Tyul'bashev V.~S., Malofeev V.~M., et al., 2018, ARep, 62, 63. doi:10.1134/S1063772918010079

\bibitem [Malofeev(2000)]{Malofeev2000} Malofeev V.~M., Malov O.~I., Shchegoleva N.~V., 2000, ARep, 44, 436. doi:10.1134/1.163868

\bibitem [Lorimer(2004)]{Lorimer2004} Lorimer D.~R.,  Kramer M., 2004, Handbook of Pulsar Astronomy, 4

\bibitem [Malofeev(2018)]{Malofeev2018} Malofeev V.~M., Tyul'bashev S.~A., 2018, RAA, 18, 096. doi:10.1088/1674-4527/18/8/96

\end{thebibliography}
\end{document}